\documentclass[conference]{IEEEtran}
\IEEEoverridecommandlockouts
\usepackage{cite}
\usepackage{amsmath,amssymb,amsfonts}
\usepackage{algorithmic}
\usepackage{graphicx}
\usepackage{textcomp}
\usepackage{xcolor}
\usepackage{booktabs}  
\usepackage{multirow}  
\usepackage{cite}
\usepackage{amsmath,amssymb,amsfonts}
\usepackage{algorithmic}
\usepackage{graphicx}
\usepackage{pifont}
\usepackage{textcomp}
\usepackage{xcolor}
\usepackage{pdfpages}
\usepackage{wrapfig}
\usepackage{float}
\usepackage{stmaryrd}

\def\BibTeX{{\rm B\kern-.05em{\sc i\kern-.025em b}\kern-.08em
    T\kern-.1667em\lower.7ex\hbox{E}\kern-.125emX}}
\begin{document}


\title{RealMind: Advancing Visual Decoding and Language Interaction via EEG Signals}



\author{\IEEEauthorblockN{Dongyang Li\textsuperscript{1,*}, Haoyang Qin\textsuperscript{1,*}, Mingyang Wu\textsuperscript{1}, Jiahua Tang\textsuperscript{3}, Yuang Cao\textsuperscript{1}, Chen Wei\textsuperscript{1,2,\dag}, Quanying Liu\textsuperscript{1,\dag}
\thanks{*: Co-first: Dongyang Li and Haoyang Qin.}}
\IEEEauthorblockA{\textsuperscript{1}Department of Biomedical Engineering, Southern University of Science and Technology, Shenzhen, China}
\IEEEauthorblockA{\textsuperscript{2}Department of Psychology, University of Birmingham, Birmingham, United Kingdom}
\IEEEauthorblockA{\textsuperscript{3}School of Computer Science and Technology, Harbin Institute of Technology, Weihai, China}
\thanks{\textsuperscript{\dag}: Corresponding to Quanying Liu and Chen Wei.}

}

\maketitle

\begin{abstract}
Decoding visual stimuli from neural recordings is a critical challenge in the development of brain-computer interfaces (BCIs). 
Although recent EEG-based decoding approaches have made progress in tasks such as visual classification, retrieval, and reconstruction, they remain constrained by unstable representation learning and a lack of interpretability. This gap highlights the need for more efficient representation learning and the integration of effective language interaction to enhance both understanding and practical usability in visual decoding tasks.
To address this limitation, we introduce RealMind, a novel EEG-based framework designed to handle a diverse range of downstream tasks. Specifically, RealMind leverages both semantic and geometric consistency learning to enhance feature representation and improve alignment across tasks. 
Notably, beyond excelling in traditional tasks, our framework marks the first attempt at visual captioning from EEG data through vision-language model (VLM). It achieves a Top-1 decoding accuracy of 27.58\% in a 200-class zero-shot retrieval task and a BLEU-1 score of 26.59\% in a 200-class zero-shot captioning task. Overall, RealMind provides a comprehensive multitask EEG decoding framework, establishing a foundational approach for EEG-based visual decoding in real-world applications.

\end{abstract}


\section{Introduction}
A key technical challenge in brain-computer interfaces (BCIs) is to decode/reconstruct the visual world seen by humans through non-invasive brain recordings, such as fMRI, MEG or EEG. EEG has long been considered incomparable to fMRI in natural image decoding/reconstruction tasks, as EEG suffers from low signal-to-noise ratio, low spatial resolution, and large inter-subject variability. Recent advances in representation alignment have made MEG/EEG visual decoding possible, although the performance is still inferior to fMRI~\cite{cichy2017multivariate,song2023decoding,grootswagers2022human}.
Despite these limitations, EEG is portable, affordable, and highly temporally resolved, making it well-suited for capturing rapid changes in brain activity during dynamic visual stimuli processing and for broad BCI applications.

Recent studies have demonstrated the efficacy of vision-language models in achieving visual decoding from functional magnetic resonance imaging (fMRI) data \cite{scotti2024reconstructing, scotti2024mindeye2, fang2024alleviating, zhou2024clip}. Leveraging multi-modal models for visual decoding from electroencephalography (EEG) signals promises to push the boundaries of current methodologies, offering enhanced accessibility and scalability for real-world applications. Building on pioneering fMRI-based approaches, several works have successfully integrated advanced multi-modal architectures to reconstruct both visual stimuli and corresponding textual descriptions from brain activity \cite{shen2024neuro, ferrante2023brain}. For instance, \textit{BrainCaption} \cite{ferrante2023brain} combines fMRI data with large language models (LLMs) and latent diffusion models (LDMs) to reconstruct images and their descriptive captions from brain activity. Similarly, \textit{BrainCLIP} \cite{liu2023brainclip} leverages contrastive learning to align fMRI data with both image and text embeddings, facilitating efficient cross-modal decoding. Additionally, \textit{UMBRAE} \cite{xia2024umbrae} introduces a universal brain encoder for multi-modal alignment, and implements cross-subject training strategies to mitigate inter-subject variability. 
While these approaches highlight the potential of large multi-modal models, their application to EEG-based visual decoding remains underexplored.

\begin{figure}[t]
    \centering
    \includegraphics[width=0.46\textwidth]{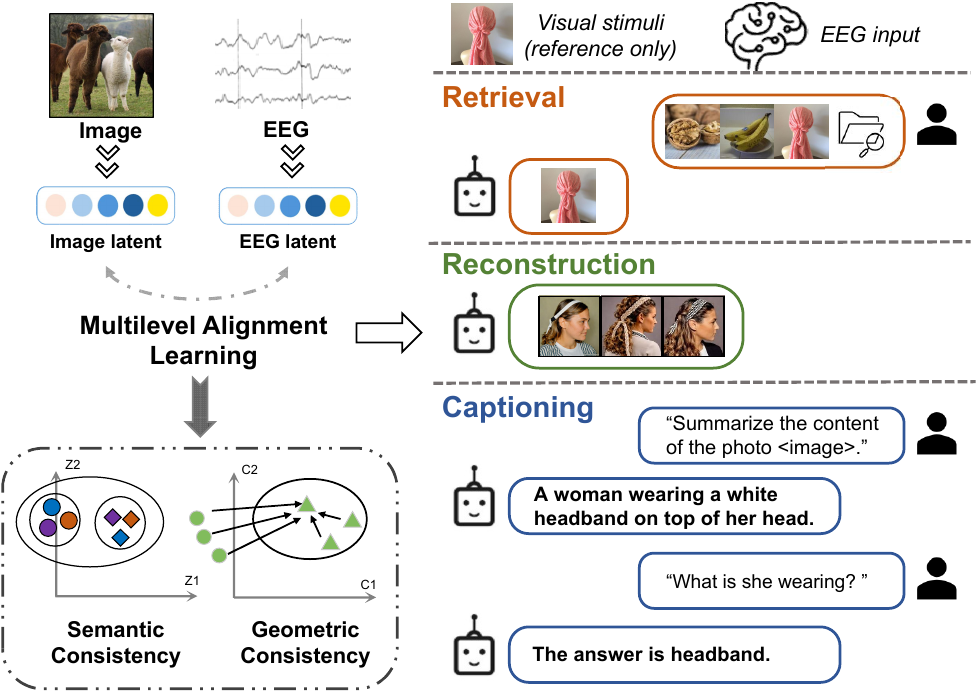}
    \caption{\textbf{Conceptual Overview}. \textbf{Left}: The \textit{RealMind} framework utilizes a multilevel representation learning strategy to align both the semantic and geometric representations of images and EEG data, incorporating constraints that enforce consistency across both modalities in terms of their underlying structure and semantics. \textbf{Right}: The aligned EEG representations facilitate the execution of a range of downstream decoding tasks, including retrieval, reconstruction, and caption generation, among others.}
    \label{fig:concep}
\end{figure}

\begin{figure*}[ht]
    \centering
    \includegraphics[page={1},trim=5.4cm 0cm 5cm 0cm, width=0.83\linewidth]{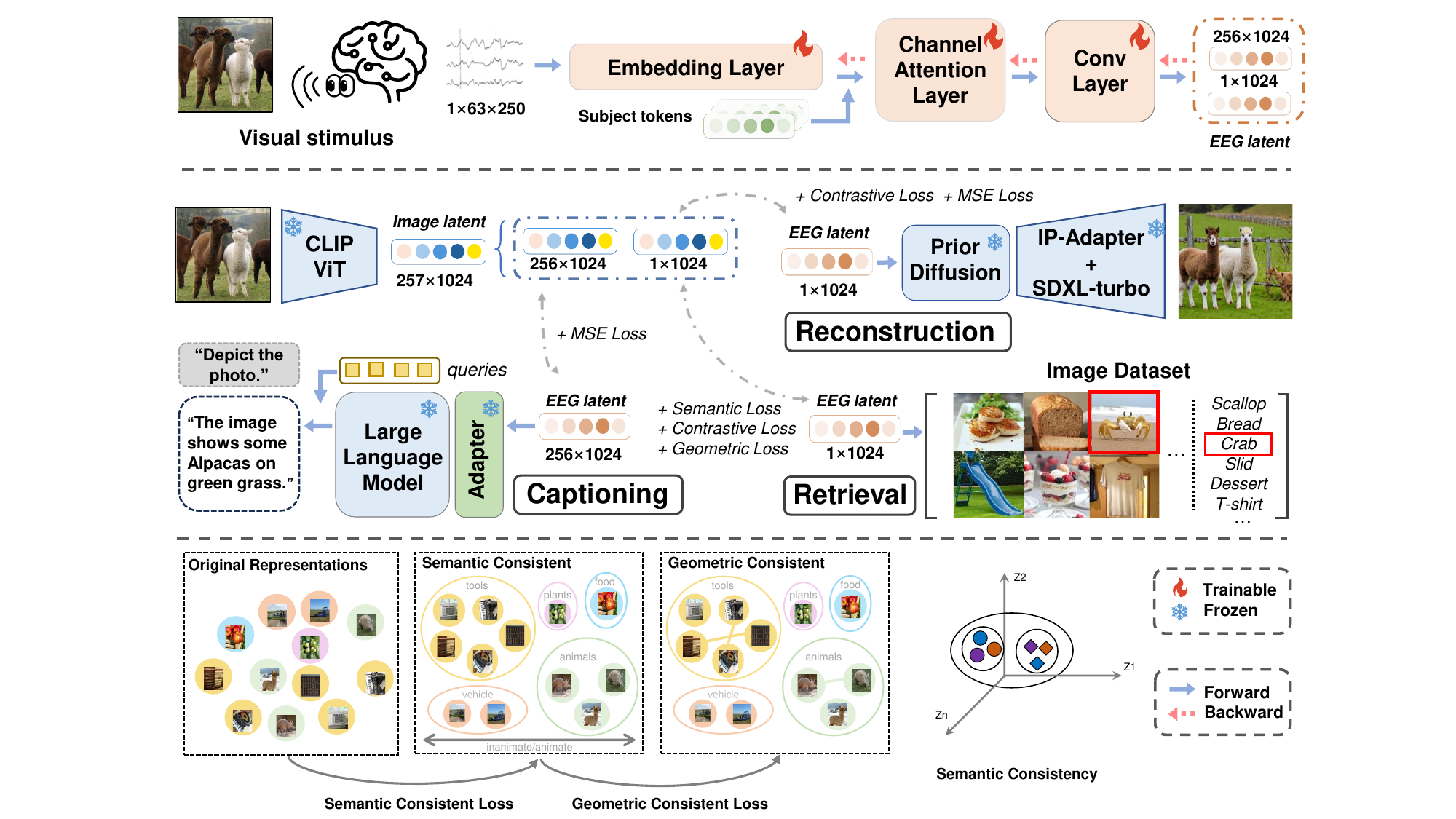}
    \caption{\textbf{RealMind framework}.
    \textbf{Top}: A Transformer-based model projects EEG signals to an latent space. \textbf{Middle}: The EEG latent with a shape of 1×1024 is aligned with the CLIP ViT-H-14 embeddings for retrieval and reconstruction of the corresponding image by SDXL. The EEG latent with a shape of 256×1024 is aligned with the CLIP ViT-L-14 embeddings to generate descriptive captions through a pre-trained large language model (LLM). \textbf{Bottom}: Brain activities that are semantically similar exhibit analogous neural patterns, and similar objects elicit comparable neural responses regardless of labels. This underscores the importance of geometric properties in compressed EEG and image representations within the feature space.}
    \label{fig:realmind}
\end{figure*}

Contrastive learning and generative models have greatly advanced EEG-based visual decoding in both decoding tasks (e.g., image classification and retrieval) and generative tasks (e.g., image reconstruction). By combining pre-trained visual models, existing EEG decoding models can learn highly-refined feature embeddings in limited data \cite{song2023decoding,li2024visual}. Using these embedded EEG features, generative models such as diffusion models can reconstruct the image one is seeing \cite{li2024visual,zhang2024cognitioncapturer}. However, research on utilizing large multi-modal models for visual captioning from EEG data remains limited \cite{palazzo2020decoding,du2023decoding,yang2024neurobind}. This is primarily due to the complexity of neural data, which is characterized by high dimensionality and substantial noise, making it challenging to develop effective EEG representations for decoding visual information—let alone generating captions merely based on EEG input.

In this study, we introduce \textbf{RealMind}  (Fig.~\ref{fig:concep}, left), designed to align EEG representations with pre-trained multi-modal models. RealMind serves as a versatile foundation for improving the performance of various downstream EEG decoding tasks (Fig.~\ref{fig:concep}, right). Our framework demonstrates significant improvements in both task-specific performance and generalization, while also offering valuable insights into cross-modal representation learning.
Our main contributions are as follows:
\begin{itemize}
    \item Through reinforced representation learning, RealMind effectively adapts to various downstream decoding tasks, including retrieval, captioning and reconstruction, supporting the practical feasibility of EEG-based visual decoding.
    \item Our results show that enforcing semantic and geometric consistency in the latent space promotes better alignment between EEG and image features, especially the achievement of a Top-5 accuracy of \textbf{58.42\%} in a 200-way retrieval task.
    \item By leveraging a pre-trained large language model, we successfully achieve, for the first time, zero-shot visual captioning from EEG data, attaining a BLEU-1 score of \textbf{26.59\%} in a 200-way captioning task.
\end{itemize}


\section{Methods}
In this section, we detail the whole framework of RealMind. We first formulate the multi-channel EEG signals as $X_i \in \mathbb{R}^{C \times T}$, where C is the number of EEG electrodes (channels) and T is the total timestamps. Let \( z_i \in \mathbb{R}^{F} \) represent the latent feature vector of image \( I_i \) and \( \hat{z}_i \) denote the latent representation of \( X_i \). Here, \( F \) corresponds to the number of features, and \( i \in \{1, 2, \dots, N\} \), with \( N \) representing the total number of images. We adapt the architecture of \cite{li2024visual}, denoted as $f_\theta$, to learn the projection from an EEG window $X_i \in \mathbb{R}^{C \times T}$ to a latent image representation $z_i \in \mathbb{R}^F$.

\subsection{Model architecture}
The original EEG sequences $X$ are first embedded into token representations independently. Channel-wise attention is then applied to these embedded tokens, enhancing interpretability by revealing correlations across electrodes. Subsequently, the representations of each token are processed by a shared feedforward network (FFN) to extract relevant features. To further improve the model's robustness and prevent overfitting, Temporal-Spatial convolution is employed, boosting the model's capability of capturing complex temporal and spatial dependencies within the EEG data. After passing through all these modules, the EEG is transformed into \( \hat{z}_i \) aligned with the image feature \( z_i \in \mathbb{R}^{F} \) extracted by CLIP. This alignment facilitates the model's application to address a range of downstream tasks.

\subsection{Training objectives}

Previous methods \cite{benchetrit2023brain, li2024visual} typically rely on contrastive and mean squared error (MSE) losses to guide the learning process, with a primary focus on mean variance and numerical similarity to adjust EEG features. These methods generally aim to increase the distance between positive and negative samples in the feature space. However, within a given training batch, stimuli from the same category but different images may be present, which can lead to the separation of stimuli even from the same category. 

To address this problem, our method introduces additional semantic consistency loss and geometric consistency loss to regulate the representation learning between EEG and image features (as shown in Figure~\ref{fig:realmind}). We train $f_\theta$ using for each batch of sampled EEG and image pairs, we compute the semantic consistency loss in the semantic space to ensure consistent alignment of cross-modal representations and the geometric consistency loss to prioritize intra-class similarity to optimize the gradient descent direction.

The objective of the semantic consistency loss can be formulated as follows:

\begin{equation}
\mathcal{L}_{Semantic}(\theta) = \frac{\left\|\mathbf{M}^I - \mathbf{M}^X\right\|_F^2}{B^2}
\end{equation}

In this equation, \(\mathbf{M}^I \in \mathbb{R}^{B \times B}\) represents the cosine similarity matrix of the image features extracted by CLIP, where $i, j \in \llbracket 1, B \rrbracket$ and each element \(\mathbf{M}^I[i, j]\) corresponds to the cosine similarity between the \(z_i\) and \(z_j\) within a given batch. Similarly, \(\mathbf{M}^X \in \mathbb{R}^{B \times B}\) represents the cosine similarity matrix of the EEG features \( \hat{z}_i \), where \(\mathbf{M}^X[i, j]\) indicates the cosine similarity between the \(\hat{z}_i\) and \(\hat{z}_j\) in the batch. Here, \(B\) denotes the batch size, and \(\|\cdot\|_F\) represents the Frobenius norm, which is employed to measure the difference between \(\mathbf{M}^I\) and \(\mathbf{M}^X\).


To achieve the geometric consistency, we first define the Gaussian potential energy between an EEG feature and an image feature:



\begin{equation}
\mathcal{V}_{Energy} = \exp \left( - \frac{\| \hat{z}_i - z_k \|^2}{2 \sigma^2} \right)
\end{equation}where \( \| \hat{z}_i - z_k \| \) is the Euclidean distance between the EEG feature \( \hat{z}_i \) and the image feature $z_k$, and \( \sigma \) is the standard deviation of the Gaussian kernel, controlling the decay rate of the potential energy.

To ensure intra-class consistency during training, for each EEG feature \( \hat{z}_i \), we randomly select \( n \) images from the same class and calculate the Gaussian potential energy between \( \hat{z}_i \) and each selected image feature $z_k$, where \( k \in \{1, 2, \dots, n\} \), where $n \in \llbracket 1, 10 \rrbracket$. Then, for the \( i \)-th EEG feature \( \hat{z}_i \), the geometric consistency loss is defined as:

\begin{equation}
\mathcal{L}_{Geometric} = \frac{1}{n} \sum_{k=1}^{n} - \mathcal{V}_{Energy} )
\end{equation}

The CLIP loss \cite{radford2021learning} is used on batches of size N with exactly one positive example:
\begin{equation}
{\mathcal{L}}_{Contrastive}(\theta)={\mathcal{L}}_{CLIP}.
\end{equation}
Next, to go beyond retrieval and instead generate and caption images, we train $f_\theta$ to directly predict the latent representations $\hat z_i$ such that we can use them to generative vision-language models. This process is performed using the standard MSE loss function over the $z_i$ and $\hat z_i$:
\begin{equation}
\mathcal{L}_{MSE}(\theta) = \frac{1}{N} \sum_{i=1}^{N} \| z_i - \hat{z}_i \|_2^2.
\end{equation}
Finally, we combine all the losses using a convex combination with tuned weight to train models that benefit from complementary training objectives. The final learning objective can be written as:
\begin{equation}
\begin{aligned}
&\mathcal{L}_{Total} = \alpha_1 \mathcal{L}_{MSE} + \\
\alpha_2 \mathcal{L}_{Contrastive} & + \alpha_3 \mathcal{L}_{Semantic} + \alpha_4 \mathcal{L}_{Geometric}
\end{aligned}
\end{equation}where the coefficients \(\alpha_1\), \(\alpha_2\), \(\alpha_3\), and \(\alpha_4\) respectively control the contributions of each loss term. 

\begin{figure*}[ht]
    \centering    
    \includegraphics[width=0.93\textwidth]{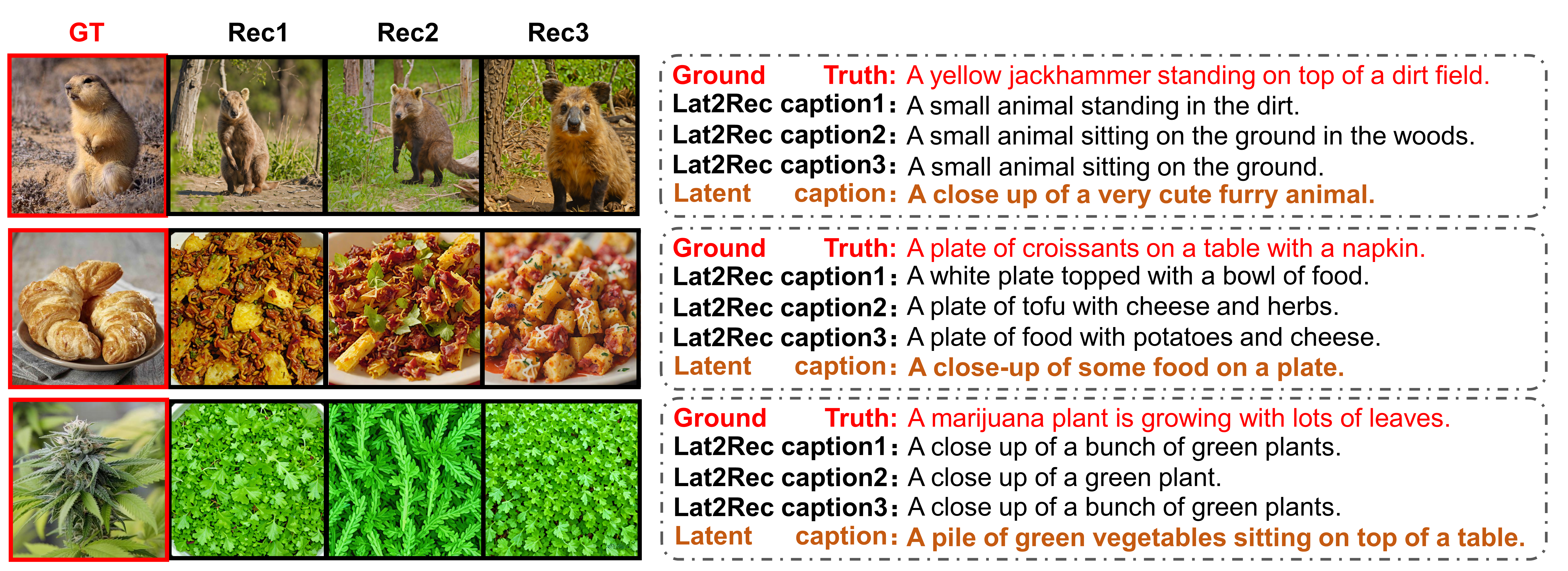}
    \caption{\textbf{Samples of EEG-based image captions generated by the RealMind framework}. We present three examples of images from subject-08. \textbf{Left}: From left to right, the original image is followed by three reconstructed images. \textbf{Right}: From top to bottom, the captions are shown for the original image (i.e., ground truth), the three reconstructed images (i.e., Lat2Rec caption), and the EEG latent representation (i.e., Latent caption).}
    \label{fig:captions}
\end{figure*}

\section{Experiments and Results}

\subsection{Experiments Setup}
We conducted experiments on the THINGS-EEG dataset \cite{gifford2022large}, which includes 1,654 distinct categories for training. Each category contains 10 images, and each image is repeated 4 times, resulting in a total of 66,160 EEG samples. These images span a broad range of visual concepts, ensuring diversity in the stimuli presented to participants. The test set comprises 200 categories, each represented by a single image repeated 80 times, yielding a total of 16,000 EEG samples. Notably, all categories in the test set are entirely distinct from those in the training set, ensuring a zero-shot evaluation scenario.

All experiments were conducted on a single NVIDIA A100 GPU. For in-subject model training, we utilized the AdamW optimizer on a dataset of 66,160 EEG samples, with an initial learning rate of $3 \times 10^{-4}$ and a batch size of 2048. Subject tokens, represented as one-hot encodings, were used to capture and distinguish individual subject characteristics. After processing through our encoder, EEG signals were transformed into a  257×1024 latent. Depending on the downstream task, a 1×1024 latent was selected for alignment with image features extracted by CLIP-ViT-H-14 for captioning tasks, while the remaining 256×1024 latent was utilized for alignment with image features extracted by CLIP-ViT-L-14 for retrieval and reconstruction tasks. To improve the signal-to-noise ratio and enhance the quality of the EEG data, we averaged the 80 repetitions for each image in the test set. This averaging process reduces the variability and noise inherent in individual EEG recordings, resulting in cleaner and more reliable data. For different downstream tasks, the choice of loss functions varies accordingly. For retrieval tasks, $\mathcal{L}_{Contrastive}, \mathcal{L}_{Semantic}, \mathcal{L}_{Geometric}$ are utilized, while for reconstruction, $\mathcal{L}_{Contrastive}, \mathcal{L}_{MSE}$ are employed and for captioning, only $\mathcal{L}_{MSE}$ are applied.

\subsection{Results on Retrieval}
To assess the efficacy of our EEG embedding approach for image retrieval, we computed the cosine similarity between the EEG-derived embeddings and the corresponding CLIP embeddings, identifying the EEG category that maximized this similarity. To ensure a comprehensive comparison, we assessed the image retrieval performance of RealMind against several alternative methods, averaging the results across all subjects in the test set. The detailed evaluation results are presented in Table~\ref{tab:accuracy}, which delineates the retrieval accuracy of different models. Compared to existing approaches, our model demonstrates a significant improvement in decoding accuracy across all retrieval tasks.

\begin{table}[h!]
\renewcommand{\arraystretch}{1.5} 
\centering
\caption{Overall accuracy of zero-shot in-subject retrieval across all subjects on THINGS-EEG dataset. We compared the 2-way, 10-way, the Top-1 and
 Top-5 accuracy of 200-way from different EEG embedding methods. }
\label{tab:accuracy}
\begin{tabular}{lccccc}
\hline
\multirow{2}{*}{Dataset} & \multirow{2}{*}{Model} & \multicolumn{1}{c}{2-way} & \multicolumn{1}{c}{10-way} & \multicolumn{2}{c}{200-way} \\
\cline{3-6}
& & Top-1 & Top-1 & Top-1 & Top-5 \\
\hline
& Benchetrit et al. \cite{benchetrit2023brain} & 91.14 & 62.62 & 16.29 & 42.16 \\
& Song et al. \cite{song2023decoding} 
& 92.17 & 65.65 & 19.40 & 46.26 \\ 
THINGS & Lawhern et al. \cite{lawhern2018eegnet} & 89.03 & 56.77 & 13.29 & 35.5 \\
& Li et al. \cite{li2024visual} & 94.60 & 73.89 & 26.09 & 58.07 \\
& \textbf{Ours} & \textbf{95.15} & \textbf{75.46} & \textbf{27.58} & \textbf{58.42} \\
\hline
\end{tabular}
\end{table}

To further evaluate the effectiveness of the two loss terms incorporated in this framework, we conducted additional experiments to assess the impact of each loss function on the overall performance. Specifically, we performed ablation studies to analyze how each loss function contributes to the experimental outcomes. In these tests, we independently removed the geometric consistency loss or semantic consistency loss, retraining the model under identical hyperparameter settings and then evaluated its retrieval performance. Since the EEG features have been aligned with the image features extracted by CLIP, the retrieval performance based solely on the EEG features provides a direct and intuitive demonstration of the impact of the remained loss function. The results of these experiments are presented in Table \ref{tab:ablation_results}, which highlights the effects of each loss function on performance metrics. 

\begin{table}[h!]
\renewcommand{\arraystretch}{1.4} 
\centering
\caption{Zero-shot in-subject retrieval ablation experiments across all subjects on the \textbf{THINGS-EEG} dataset assess Contrastive Loss (CL), Semantic Loss (SL), and Geometric Loss (GL).}
\label{tab:ablation_results}
\resizebox{\columnwidth}{!}{
\begin{tabular}{ccc ccc cc}
\hline
\multirow{2}{*}{\centering \textbf{CL}} & \multirow{2}{*}{\centering \textbf{SL}} & \multirow{2}{*}{\centering \textbf{GL}} & \textbf{2-way} & \textbf{10-way} & \multicolumn{2}{c}{\textbf{200-way}} \\ 
\cline{4-7}
 & & & \textbf{top-1} & \textbf{top-1} & \textbf{top-1} & \textbf{top-5} \\ 
\hline
\checkmark & \ding{55} & \ding{55} & 94.63 & 73.04 & 24.19 & 55.64 \\
\checkmark & \checkmark & \ding{55} & 94.56 & 74.60 & 25.59 & 57.19 \\
\checkmark & \ding{55} & \checkmark & 94.95 & 74.67 & 25.63 & 57.59 \\
\checkmark & \checkmark & \checkmark & \textbf{95.15} & \textbf{75.46} & \textbf{27.58} & \textbf{58.42} \\
\hline
\end{tabular}
}
\end{table}

\subsection{Results on Reconstruction}
To highlight the versatility of our method, we conducted experiments on visual reconstruction task following the method outlined in \cite{li2024visual}. We evaluated and compared the reconstruction performance for EEG, MEG, and fMRI across various metrics from several advanced methods and datasets. The results are summarized in the Table~\ref{tab:generation_comparison}. 


\begin{table}[h!]
\renewcommand{\arraystretch}{1.5} 
\centering
\caption{Quantitative assessments of the reconstruction quality for EEG, MEG, and fMRI across different methods in Subject 8.}
\label{tab:generation_comparison}
\resizebox{\columnwidth}{!}{
\begin{tabular}{lcccc} 
\toprule
{Dataset} & \multicolumn{2}{c}{Low-level} & \multicolumn{2}{c}{High-level} \\
\cmidrule(r){2-3} \cmidrule(r){4-5}
   & SSIM $\uparrow$ & AlexNet(2) $\uparrow$ & AlexNet(5) $\uparrow$ & CLIP $\uparrow$ \\
\midrule
B.D./fMRI \cite{Benchetrit_2023} & 0.366 & 0.962 & 0.977 & 0.917 \\
BrainDiffuser/fMRI~\cite{ozcelik2023natural} & 0.356  & 0.942  & 0.962 & 0.915 \\
MindEye/fMRI \cite{scotti2024reconstructing} & 0.308  & 0.917  & 0.974 & 0.942 \\ \hline
ATM/MEG~\cite{li2024visual} & 0.340  & 0.613  & 0.672 & 0.603 \\
ATM/EEG~\cite{li2024visual} & 0.345 & 0.776  & 0.866 & 0.786 \\
\textbf{Ours/EEG} & \textbf{0.373} & \textbf{0.797}  & \textbf{0.884} & \textbf{0.813} \\
\bottomrule
\end{tabular}
}
\end{table}


\subsection{Results on Captioning}

\begin{figure}[h!]
    \centering
    \includegraphics[page=1, trim=9.7cm 6.2cm 10cm 5.45cm, clip,width=0.5\textwidth]{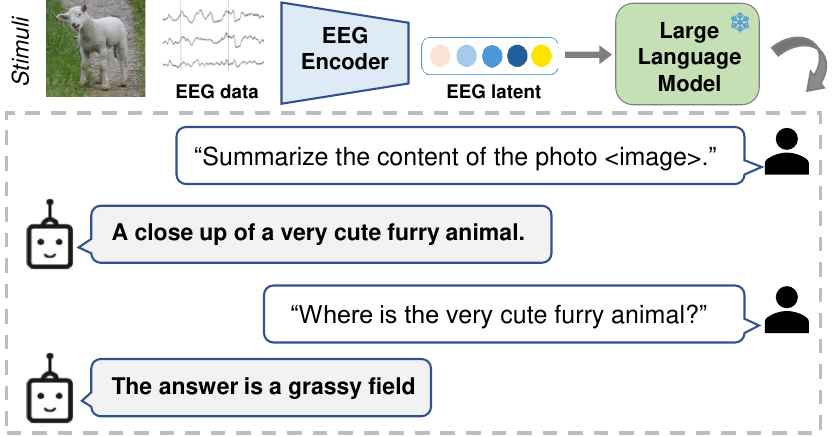}
    \caption{\textbf{Examples of generated answers using RealMind}. Different task prompts for the same input brain signal result in unique outcomes.}
    \label{fig:QA}
\end{figure}

We equipped RealMind EEG representations with the Shikra model for captioning task (in Fig.~\ref{fig:captions}) and employed the two-stage reconstruction method proposed by \cite{li2024visual} to generate reconstructed images. On the left side of Fig.~\ref{fig:captions}, we present the images reconstructed from EEG data, while the right side shows the corresponding captions for these images. The captions were generated using two different approaches: (1) Latent caption: For a given EEG signal, a caption can be directly generated from the extracted EEG features. (2) Lat2Rec caption: an image is first reconstructed from the EEG data, followed with caption generation from the reconstructed image. Additionally, the extracted EEG features can be directly used for Visual Question Answering (VQA) task (in Fig.~\ref{fig:QA}). 

\begin{table}[h]
\renewcommand{\arraystretch}{1.5} 
\centering
\caption{ The above method was trained and tested in the NSD dataset, and our method was trained and tested using the THINGS-EEG dataset in Subject 8.}
\label{tab:caption_effect_between_decoders}
\scalebox{0.9}{
\begin{tabular}{lccccc}
\hline
\textbf{Method} & \textbf{BLEU1} & \textbf{BLEU4} & \textbf{METEOR} & \textbf{Sentence} & \textbf{CLIP} \\
\hline
Shikra-w/image~\cite{chen2023shikra} & 82.38 & 49.66 & 35.60 & 65.49 & 80.60 \\
\hline
SDRecon/fMRI~\cite{takagi2023high} & 36.21 & 3.43 & 10.03 & 25.13 & 61.07 \\
OneLLM/fMRI~\cite{han2024onellm} & 47.04 & 9.51 & 13.55 & 35.05 & 54.80 \\
BrainCap/fMRI~\cite{ferrante2023brain} & 55.96 & 14.51 & 16.68 & 40.69 & 64.31 \\
UMBRAE/fMRI~\cite{xia2024umbrae} & 57.84 & 17.17 & 18.70 & 42.14 & 66.10 \\
\hline
\textbf{Ours/EEG} & \textbf{26.59} & \textbf{4.31} & \textbf{17.79} & \textbf{17.76} & \textbf{55.78} \\
\hline
\end{tabular}
}
\end{table}

The results demonstrate that RealMind effectively captures and represents the visual stimuli experienced by subjects throughout the experimental process. To the best of our knowledge, this study represents the first effort to apply the image captioning task to EEG data, and the results reveal that RealMind achieves competitive performance in this context, even when compared to fMRI, as shown in Table \ref{tab:caption_effect_between_decoders}.


In zero-shot scenarios, our approach outperforms methods that first reconstruct the image and then generate captions, as shown in Table~\ref{tab:eeg_image_caption} and Fig.~\ref{fig:captions}. The approach of generating captions directly from EEG features benefits from a robust alignment between the EEG features and the visual representations during the representation learning phase. This alignment prevents information loss during the reconstruction process, thereby ensuring that the generated captions are more accurate.

\begin{table}[h!]
\centering
\renewcommand{\arraystretch}{1.5}  \
\caption{Evaluation of EEG-to-image captions in Subject 8. We used different ground-truth captions for comparison (e.g., Shikra \cite{chen2023shikra} captions, GIT \cite{wang2022git} captions or captions generated from BLIP2 \cite{li2023blip}). \textit{Abbrev.} L2Cap: EEG latent to caption; I2Cap: image to caption (i.e., reconstructing images from EEG and then generating captions).}
\label{tab:eeg_image_caption}

\begin{tabular}{lcccccc}
\hline
\multirow{2}{*}{Metric} & \multicolumn{2}{c}{Shikra captions} & \multicolumn{2}{c}{GIT captions} & \multicolumn{2}{c}{BLIP2 captions} \\
\cline{2-7}
 & L2Cap & I2Cap & L2Cap & I2Cap & L2Cap & I2Cap \\
 
\hline
BLEU-1 $\uparrow$      & \textbf{26.59} & 23.09 & 15.43 & 18.28 & 18.31 & 25.97 \\
BLEU-4 $\uparrow$      & \textbf{4.31}  & 3.88  & 2.90  & 3.70  & 3.25  & 4.65 \\
METEOR $\downarrow$    & \textbf{17.79} & 15.00 & 15.43 & 14.20 & 15.01 & 18.40 \\
Sentence $\uparrow$    & \textbf{17.76} & 19.62 & 14.26 & 23.78 & 15.60 & 25.99 \\
CLIP-ViT-L $\uparrow$  & \textbf{55.78} & 53.91 & 57.83 & 61.34 & 58.77 & 57.52 \\
\hline
\end{tabular}

\end{table}


\section{Discussion and Conclusion}
In this paper, we introduced RealMind, a novel framework for EEG-based visual decoding. By integrating semantic and geometric constraints during the representation learning phase, RealMind achieved more efficient alignment with human cognitive processes, significantly enhancing the accuracy and robustness of EEG-based visual semantic decoding. Our model delivered state-of-the-art performance in both retrieval and reconstruction tasks. Additionally, by incorporating a large language model, RealMind successfully generated credible captions from EEG representations. To our knowledge, RealMind is the first framework to achieve zero-shot visual captioning using EEG data.

RealMind advances current EEG visual decoding approaches in two key aspects. First, we introduced a loss function specifically designed for EEG data, significantly improving alignment performance. EEG, compared to fMRI, has long faced challenges such as high noise levels and poor data quality, which have hindered its decoding performance \cite{benchetrit2023brain}. Our results demonstrate that the additional semantic and geometric loss functions enhanced data utilization efficiency, leading to improved decoding performance. This is attributed to better feature alignment and the successful integration of large language models. Moreover, we presented the first effective EEG-based captioning solution, paving the way for practical applications in EEG decoding. Compared to fMRI-based visual decoding methods, EEG offers a cost-effective and more accessible alternative. Here, we further extend EEG-based visual decoding to zero-shot visual captioning tasks.

In the future, we aim to deepen the integration of EEG with multimodal data by incorporating neural data with other data types into a unified multimodal model, rather than simply adding isolated modules. Aligned with the trend of multimodal models, our goal is to develop a single, unified framework capable of handling multiple EEG-based decoding tasks, such as retrieval, reconstruction, and captioning. This approach will enhance data utilization efficiency, enabling limited EEG data to more effectively leverage information from visual and language modalities. Ultimately, this will provide a strong foundation for the broader application of EEG-based BCI systems.






\bibliographystyle{IEEEbib}
\bibliography{icme2025references}


\end{document}